\documentclass{article}
\usepackage{tikz}
\usepackage{pgf}
\usepackage[T1]{fontenc}
\usepackage[utf8]{inputenc}
\usepackage[english]{babel}
\usepackage{graphicx}
\usepackage{amsmath,amsfonts,amsthm,amsopn,mathtools}
\usepackage{floatflt}
\usepackage{amssymb}
\usepackage{xfrac}
\usepackage{listings}
\usepackage{color}
\usepackage{authblk}
\usepackage{subfigure}

\newcommand{\vel}{\mathbf{u}}
\newcommand{\divtau}{\nabla \cdot \boldsymbol{\tau}}
\newcommand{\tr}{\textnormal{tr}}
\newcommand{\btau}{\boldsymbol{\tau}}
\newcommand{\divs}{\widetilde{\nabla \cdot \mathbf{S}}}
\newcommand{\gradk}{\widetilde{\nabla k}}
\newcommand{\RANS}{\textnormal{RANS}}
\newcommand{\DNS}{\textnormal{DNS}}
\newcommand{\NN}{\textnormal{NN}}

\definecolor{mygreen}{rgb}{0, 0.6, 0} 

\usepackage{xspace}
\newcommand{\openfoam}{Open\nolinebreak\hspace{-.2em}{\color{blue}\Large$\nabla$}\nolinebreak\hspace{-.2em}FOAM\textsuperscript{\textregistered}\xspace}

\title{A data-driven approach for the closure of RANS models by the divergence of the Reynolds Stress Tensor}

\author[$\dag$]{S. BERRONE}
\author[$\dag$]{D. OBERTO \thanks{Corresponding author: davide.oberto@polito.it}}

\affil[$\dag$]{Dipartimento di Scienze Matematiche, Politecnico di Torino, Corso Duca degli Abruzzi 24, 10129 Torino, Italy.}

\providecommand{\keywords}[1]{\textbf{\textit{Keywords  }} #1}

\begin{document}
\maketitle

\begin{abstract}
In the present paper a new data-driven model is proposed to close and increase accuracy of RANS equations. The divergence of the Reynolds Stress Tensor (RST) is obtained through a Neural Network (NN) whose architecture and input choice guarantee both Galilean and coordinates-frame rotation. The former derives from the input choice of the NN while the latter from the expansion of the divergence of the RST into a vector basis. This approach has been widely used for data-driven models for the anisotropic RST or the RST discrepancies and it is here proposed for the divergence of the RST. Hence, a constitutive relation of the divergence of the RST from mean quantities is proposed to obtain such expansion. Moreover, once the proposed data-driven approach is trained, there is no need to run any classic turbulence model to close the equations. 

The well-known tests of flow in a square duct and over periodic hills are used to show advantages of the present method compared to standard turbulence models. 
\end{abstract}


\vskip2.0cm

%
%

\section{Introduction} \label{Sec:Intro}
Reynolds-Averaged Navier-Stokes (RANS) equations are widely used in engineering for turbulent flow simulations. Their popularity comes from the low computational cost compared to Large-Eddy Simulations (LES) and Direct Numerical Simulation (DNS) approaches. However, RANS predictions may be highly inaccurate for some classes of flows \cite{Craft1996} due to the bad description of the effects of the Reynolds stresses on the mean flow \cite{Oliver2011}. On the other hand, thanks to the remarkable growth of HPC facilities, more and more DNS data coming from simple geometries and moderate Reynolds numbers are becoming available, even if usually only partially as observed in \cite{Xiao2020}. Some classes of flows with DNS datasets are: channel flows \cite{Moser1999,Abe2001}, pipe and duct flows \cite{Pirozzoli2021,Pinelli2010,Zhang2015,Pirozzoli2018}, flows over periodic hills \cite{Breuer2009,Xiao2020}, flows around cylinders \cite{Trias2015,Cimarelli2018,Chiarini2021}. As a consequence, in the past years an increasing number of studies took advantage of Machine Learning techniques to exploit DNS data. The remarkable variety of review articles in the recent years on the subject highlights this trend \cite{Kutz2017,Duraisamy2019,Brunton2020,Vinuesa2021a}. In particular, one active research area is focused on data-driven RANS turbulence models that increase accuracy through DNS (or highly accurate LES) data.

In this framework, data-driven models must satisfy the same invariance properties of the physical system they are modelling. In \cite{Ling2016a} it is proved that invariance to coordinates-frame rotation can be guaranteed by taking for each physical dimension 10 rotations of the initial dataset and by including them in an augmented dataset. Even if this approach is conceptually valid, it has the huge drawback that for 3D problems the dataset would become 1000 times the initial one making this method impracticable both in terms of learning time and of memory storage.

Another approach to guarantee invariance properties has been proposed in the pioneering paper \cite{Ling2016}. In this work, invariance is automatically satisfied by the architecture of the trained Neural Network (NN). In particular, the NN outputs were the coefficients of the decomposition of the anisotropic Reynolds Stress Tensor into a tensor basis defined in \cite{Pope1975}, while the inputs were invariant quantities. The obtained neural network was named Tensor Basis Neural Network (TBNN). Since then, various studies have been performed to analyse data-driven approaches for the anisotropic Reynolds Stress Tensor. In \cite{Fang2019} a change to the TBNN was proposed to increase accuracy. In \cite{OcarizBorde2021} a Convolutional Neural Network was used for 1D turbulent flows and its interpretability was discussed. In \cite{Vinuesa2021} both the coefficients and the basis of the anisotropy Reynolds stress tensor were inferred. In \cite{Steiner2022} this approach was applied for wind turbine wakes under neutral conditions.
The same rotation-invariance idea was used in \cite{Wang2017,Wu2018} to train random forests able to predict the discrepancy between the Reynolds Stress Tensor (RST) obtained by a baseline RANS turbulence model and the DNS one. 

In \cite{Thompson2019} it was shown that predicting the divergence of the Reynolds Stress Tensor, denoted as Reynolds Force Vector (RFV), instead of the RST itself can effectively increase the accuracy of data-driven RANS turbulence models. From one hand, the RFV can be obtained from first order statistics reducing intrinsic statistical errors of DNS data. On the other hand, the RFV directly compares into the RANS equations and it seems natural to directly predict it. 
This work is grounded on \cite{Thompson2019} and aims to enforce into the RANS system physical invariance properties by construction without data augmentation. Analogously to \cite{Ling2016}, in this work a constitutive assumption of the RFV from mean fields is derived. This hypothesis is fundamental to derive the inputs of the data-driven model and the vector basis used to expand the RFV. The obtained neural network is called Vector Basis Neural Network (VBNN).
The proposed approach closes the RANS system without the requirements of additional PDEs for turbulent scalar quantities or for the RST discrepancies as in previous works. As a consequence, once the model is satisfactorily trained, it does not require any coupling with a classic turbulence model.

Besides this introduction, the paper is organized into four more sections. In Section 2 a brief overview on RANS models is given to successively describe the TBNN presented in \cite{Ling2016}. In Section 3 the constitutive dependencies of the divergence of the RST are derived. The properties of the VBNN are consequently discussed. The numerical results are presented and discussed in Section 4. Classic benchmark flows in a square duct and over periodic hills are chosen as numerical experiments to analyse the data-driven model for two main reasons: i) availability of DNS data in literature; ii) despite their geometrical simplicity, classic RANS turbulence models fail in the prediction of their velocity fields. Finally, in the last section conclusions are drawn.

%
%

\section{Tensor Basis Neural Network} \label{Sec:TBNN}

\subsection{Reynolds-Averaged Navier-Stokes equation and turbulence models}
The RANS equations for incompressible flows  read
\begin{equation} \label{eq:RANS}
\begin{dcases}
&\nabla \cdot \vel=0 \\
&\frac{\partial \vel}{\partial t}+\vel \cdot \nabla\vel - \nu \Delta \vel = - \nabla p - \divtau,
\end{dcases}
\end{equation}
where $\vel$ is the averaged velocity field, $\nu$ is the kinematic viscosity of the fluid, $p$ is the averaged pressure field normalized by the constant density of the fluid and $\boldsymbol{\tau}$ is the Reynolds Stress Tensor. The latter is a symmetric tensor that needs to be modelled to close the RANS equations and whose components are associated to the correlations of the turbulent fluctuating components of the velocity field. Hence, the divergence of the RST describes the effects of the turbulence on the averaged fields.

One class of turbulence models, called \emph{linear isotropic models}, is based on the well known Boussinesq hypothesis. The RST is modelled as 
\begin{equation} \label{eq:linear_closures}
\boldsymbol{\tau} = \frac{2}{3} k \mathbf{I} - 2 \nu_t \mathbf{S}
\end{equation} 
where $k=\frac{1}{2} \tr(\boldsymbol{\tau})$ is the \emph{turbulent kinetic energy} ($\tr$ denotes the trace operator), $\mathbf{I}$ is the identity tensor, $\nu_t$ is the \emph{turbulent viscosity} and $\mathbf{S} = \frac{1}{2} [\nabla \vel + (\nabla \vel)^T]$ is the mean strain rate tensor. The quantity $\nu_t$ must be modelled and the system is usually closed by two PDEs (for example one for the turbulent kinetic energy $k$ and one for its dissipation rate $\varepsilon$ using the relation $\nu_t = C_\mu k^2/\varepsilon$ being $C_\mu$ a model's constant). In literature many different linear turbulence models are defined depending, for example, on the choice of the variables solved to model $\nu_t$. 

The linear isotropic models fail in the description of some physical behaviours. As a consequence, more advanced non-linear models have been proposed in literature such as \cite{Pope1975,Craft1996}. These models assume an algebraic representation of $\boldsymbol{\tau}$ more complex than in \eqref{eq:linear_closures}. Indeed, they include dependences on high-order powers of the mean strain rate tensor and the mean rotation rate tensor $\mathbf{W} = \frac{1}{2} [\nabla \vel - (\nabla \vel)^T]$.

Another classical approach to close the RANS system \eqref{eq:RANS} is to solve a PDE for each component of the RST tensor. This class of models are called Reynolds Stress Transport Models (RSTM). This procedure does not require any modeling of the RST but, on the other hand, require the modeling of some terms inside the PDEs. Classic examples of RSTM are \cite{Launder1975,LRR,Speziale1991}.

Unfortunately, both non-linear and RSTM turbulence models are more likely to diverge compared to classic linear models and, consequently, the latter are still the main option for many flow cases.

\subsection{Tensor Basis Neural Network}
Let $\mathbf{a}:= \frac{\btau}{2 k} - \frac{1}{3} \mathbf{I}$ be the \emph{anisotropic Reynolds Stress Tensor} where $\btau$ is the Reynolds Stress Tensor, $k$ is the turbulent kinetic energy and $\mathbf{I}$ is the identity tensor. The tensor $\mathbf{a}$ is dimensionless with vanishing trace by definition. 

Let $\mathbf{s} =  \frac{1}{2} \frac{k}{\varepsilon} [\nabla \vel + (\nabla \vel)^T]$ and $\mathbf{w} = \frac{1}{2} \frac{k}{\varepsilon}  [\nabla \vel - (\nabla \vel)^T]$ be the dimensionless counterparts of the mean strain rate tensor $\mathbf{S}$ and mean rotation rate tensor $\mathbf{W}$ respectively, where $\varepsilon$ is the turbulent kinetic energy dissipation rate.

Some classic algebraic turbulence models for the RST can be rephrased as algebraic models for its anisotropic part in terms of $\mathbf{s}$ and $\mathbf{w}$. For example the classic linear closure \eqref{eq:linear_closures} is equivalent to $\mathbf{a} = - \frac{\varepsilon}{k^2} \nu_t \mathbf{s}$. 

In \cite{Pope1975} a more general constitutive relation $\mathbf{a}=\mathbf{a}(\mathbf{s},\mathbf{w})$ was supposed. This assumption and the Cayley-Hamilton theorem lead to
\begin{equation} \label{eq:aRST_expansion}
\mathbf{a} = \sum_{j=1}^{10} {c_j (\lambda_1,\dots,\lambda_6)\mathbf{T}_j},
\end{equation}
where $\lambda_i$, $i=1,\dots,6$, are invariant scalar quantities that depend on $\mathbf{s}$ and $\mathbf{w}$. Following \cite{Pope1975,Zheng1994} they are
\begin{equation} \label{eq:invariants_a}
\begin{split}
& \lambda_1 = \tr(\mathbf{s}^2), \quad \lambda_2 = \tr(\mathbf{s}^3), \quad  \lambda_3 = \tr(\mathbf{w}^2), 
\\
&\lambda_4 = \tr(\mathbf{s}\mathbf{w}^2), \quad \lambda_5 = \tr(\mathbf{s}^2\mathbf{w}^2), \quad \lambda_6 = \tr(\mathbf{s}^2\mathbf{w}^2\mathbf{s}\mathbf{w}).
\end{split}
\end{equation}
Furthermore, $\mathbf{T}_j$, $j=1, \dots, 10,$ are 
\begin{equation} \label{eq:basis_a}
\begin{split}
&\mathbf{T}_1 = \mathbf{s},  \quad \mathbf{T}_2 = \mathbf{sw}+\mathbf{ws}, \quad \mathbf{T}_3 = \mathbf{s}^2 - \frac{1}{3} \tr(\mathbf{s}^2) \mathbf{I}, \quad \mathbf{T}_4 = \mathbf{w}^2 - \frac{1}{3} \tr(\mathbf{w}^2) \mathbf{I},  
\\
& \mathbf{T}_5 = \mathbf{w}\mathbf{s}^2 - \ \mathbf{s}^2\mathbf{w}, \quad \mathbf{T}_6 = \mathbf{w}^2\mathbf{s}+\mathbf{s}\mathbf{w}^2 - \frac{2}{3} \tr(\mathbf{s}\mathbf{w}^2)\mathbf{I}, \quad \mathbf{T}_7 = \mathbf{w}\mathbf{s}\mathbf{w}^2 - \mathbf{w}^2\mathbf{s}\mathbf{w}, \\ 
&\mathbf{T}_8 = \mathbf{s}\mathbf{w}\mathbf{s}^2 - \mathbf{s}^2\mathbf{w}\mathbf{s}, \quad \mathbf{T}_9 = \mathbf{w}^2\mathbf{s}^2+\mathbf{s}^2\mathbf{w}^2 - \frac{2}{3} \tr(\mathbf{s}^2\mathbf{w}^2)\mathbf{I},\\
& \mathbf{T}_{10} = \mathbf{w}\mathbf{s}^2\mathbf{w}^2 - \mathbf{w}^2\mathbf{s}^2\mathbf{w}.
\end{split}
\end{equation}

Equation \eqref{eq:aRST_expansion} states that the anisotropic Reynolds Stress Tensor can be expressed as a finite linear combination of the 10 tensor basis elements $\{\mathbf{T}_j\}$ listed above. Moreover, the coefficients involved are functions of the 6 invariants $\{\lambda_i\}$. Both the tensor basis and the invariants are known \emph{a priori}. The only unknowns are the expressions of the 10 coefficients $\{c_j\}$.

The formula \eqref{eq:aRST_expansion} has been the starting point in \cite{Ling2016} to define a Tensor Basis Neural Network able to predict the coefficients using the invariants as inputs. Once the coefficients are obtained, the linear combination is computed to obtain $\mathbf{a}$. This approach has the huge advantage that the coefficients are automatically invariant to coordinates-frame rotations and Galilean transformations. This property arises from the Galilean and coordinates-frame rotation invariance of the inputs $\{\lambda_i\}$ that are fed into the TBNN. 

%
%

\section{Vector Basis Neural Network} \label{Sec:VBNN}
This section focuses on the Vector Basis Neural Network used in this work to close the RANS system by obtaining the $\divtau$ term.

\subsection{Constitutive dependencies}
Let us define the dimensionless quantity $\widetilde{\divtau} = \frac{k^{1/2}}{\varepsilon} \divtau$. In the present work we assume the constitutive hypothesis
\begin{equation} \label{eq:divtau_const_hp}
\widetilde{\divtau} = \mathbf{f}(\mathbf{s},\mathbf{w},\widetilde{\nabla \cdot \mathbf{S}},\widetilde{\nabla k}, Re_d),
\end{equation}
where $\mathbf{s} = \frac{k}{\varepsilon}  \mathbf{S}$, $\mathbf{w} = \frac{k}{\varepsilon}  \mathbf{W}$, $\widetilde{\nabla \cdot \mathbf{S}} = \frac{k^{5/2}}{\varepsilon^2}  \nabla \cdot \mathbf{S}$ and $\widetilde{\nabla k} = \frac{k^{1/2}}{\varepsilon}  \nabla k$ are the dimensionless counterparts of the symmetric part of the velocity gradient $\mathbf{S}$, the antisymmetric part of the velocity gradient $\mathbf{W}$, the divergence of $\mathbf{S}$ and the gradient of $k$,  respectively. Finally, $Re_d = \min(\frac{\sqrt{k} d}{50 \nu},2)$ is the wall-distance based Reynolds number, where $d$ is the wall distance. This quantity is relevant during the training process as reported in \cite{Wang2017, Milani2020}. The motivations behind this constitutive choice are discussed in Appendix. 

With the above hypothesis, following the idea in \cite{Ling2016},  $\widetilde{\divtau}$ can be written in a basis made by $N_c=12$ vectors $\{\mathbf{t}_k\}_{k=1}^{N_c}$ with corresponding coefficients that depend on $N_i=26$ invariant scalar quantities $\{\lambda_k\}_{k=1}^{N_i}$. In particular, it reads 
\begin{equation} \label{eq:divtau_expansion}
\widetilde{\divtau} = \sum_{k=1}^{N_c}{c_k (\lambda_1,\dots,\lambda_{N_i}) \ \mathbf{t}_k}.
\end{equation}

The appropriate vector basis and the invariants can be obtained from \cite{Zheng1994} (Tables 1,2). The vector basis reads
\begin{equation} \label{eq:basis_div_tau}
\begin{split}
&\mathbf{t}_1 = \divs, \quad \mathbf{t}_2 = \mathbf{s} \ \divs, \quad \mathbf{t}_3 = \mathbf{s}^2 \ \divs,
\\
& \mathbf{t}_4 = \mathbf{w} \ \divs, \quad \mathbf{t}_5 = \mathbf{w}^2 \ \divs, \quad \mathbf{t}_6 = (\mathbf{sw}+\mathbf{ws}) \ \divs, \\ 
\noalign{\vskip5pt}
&\mathbf{t}_7 = \gradk, \quad \mathbf{t}_8 = \mathbf{s} \ \gradk, \quad \mathbf{t}_9 = \mathbf{s}^2 \ \gradk,
\\
& \mathbf{t}_{10} = \mathbf{w} \ \gradk, \quad \mathbf{t}_{11} = \mathbf{w}^2 \ \gradk, \quad \mathbf{t}_{12} = (\mathbf{sw}+\mathbf{ws}) \ \gradk.
\end{split}
\end{equation}
The invariants are
\begin{equation} \label{eq:invariants_div_tau}
\begin{split}
&\lambda_1 = (\divs)^T (\divs), \quad \lambda_2 = \tr(\mathbf{s}^2), \quad \lambda_3 = \tr(\mathbf{s}^3), \quad  \lambda_4 = \tr(\mathbf{w}^2), 
\\
&\lambda_5 = \tr(\mathbf{s}\mathbf{w}^2), \quad \lambda_6 = \tr(\mathbf{s}^2\mathbf{w}^2), \quad \lambda_7 = \tr(\mathbf{s}^2\mathbf{w}^2\mathbf{s}\mathbf{w}), \quad \lambda_8 = (\divs)^T \mathbf{s} (\divs), 
\\
&\lambda_9 = (\divs)^T \mathbf{s}^2 (\divs), \quad \lambda_{10} = (\divs)^T \mathbf{w}^2 (\divs), \quad  \lambda_{11} = (\divs)^T \mathbf{s}\mathbf{w} (\divs),
\\
&\lambda_{12} = (\divs)^T \mathbf{s}^2 \mathbf{w} (\divs), \quad \lambda_{13} = (\divs)^T \mathbf{w}\mathbf{s}\mathbf{w}^2 (\divs),\\ 
\noalign{\vskip5pt}
&\lambda_{14} = (\gradk)^T (\gradk), \quad \lambda_{15} = (\gradk)^T \mathbf{s} (\gradk), \quad \lambda_{16} = (\gradk)^T \mathbf{s}^2 (\gradk), \\
& \lambda_{17} = (\gradk)^T \mathbf{w}^2 (\gradk), \quad \lambda_{18} = (\gradk)^T \ \divs,
\quad  \lambda_{19} = (\gradk)^T \mathbf{s}\mathbf{w} (\gradk),
\\
&\lambda_{20} = (\gradk)^T \mathbf{s}^2 \mathbf{w} (\gradk), \quad \lambda_{21} = (\gradk)^T \mathbf{w}\mathbf{s}\mathbf{w}^2 (\gradk), \quad \lambda_{22} = (\gradk)^T \mathbf{s} (\divs), \\
&\lambda_{23} = (\gradk)^T \mathbf{s}^2 (\divs), \quad \lambda_{24} = (\gradk)^T  \mathbf{w} (\divs), \quad
\lambda_{25} = (\gradk)^T \mathbf{w}^2 (\divs), \\ 
&\lambda_{26} = (\gradk)^T (\mathbf{s}\mathbf{w} + \mathbf{w}\mathbf{s}) (\divs) , \quad \lambda_{27} = Re_d,
\end{split}
\end{equation}
where the first 26 invariants derive from the dependencies on $\mathbf{s},\mathbf{w},\widetilde{\nabla \cdot \mathbf{S}},\widetilde{\nabla k}$ while the last one is the scalar quantity (and consequently invariant to the choice of the coordinates-frame) added in the dependencies assumption \eqref{eq:divtau_const_hp}. The invariant $\tr(\mathbf{s})$ is neglected because identically zero due to the incompressibility constraint.

In Section \ref{Sec:Results}, we will consider also the simplified assumption 
\begin{equation*}
\widetilde{\divtau} = \mathbf{f}(\mathbf{s},\mathbf{w},\widetilde{\nabla \cdot \mathbf{S}}, Re_d).
\end{equation*}
In this case, the vector basis is formed by the first 6 vectors in \eqref{eq:basis_div_tau} while the invariants are the first 13 and the last one in \eqref{eq:invariants_div_tau} (they are the expressions that do not involve $\gradk$).

\subsection{Vector Basis Neural Network}

\subsubsection{Inputs and outputs}
The Vector Basis Neural Network obtains the coefficients $c_j$, $j=1, \dots, N_c$, in \eqref{eq:divtau_expansion} to be multiplied to the vector basis elements. The VBNN should be able to reproduce the divergence of the RST using only informations coming from RANS simulations. In particular, during the training stage, the quantity $\divtau$ comes from the DNS while the invariants, the vectors and the adimensionalization factors come from the RANS. Therefore, during the training the optimization process aims to reduce
\begin{equation} \label{eq:to_minimize}
\vert \vert \ (\widetilde{\divtau})^\DNS - \sum_{k=1}^{N_c}{c^\NN_k (\lambda^\RANS_1,\dots,\lambda^\RANS_{N_i}) \ \mathbf{t}^\RANS_k} \ \vert \vert_2.
\end{equation}
Here, the quantities $\lambda^\RANS_i$, $i=1,\dots,N_i$, and $\mathbf{t}^\RANS_k$, $k = 1,\dots,N_c$, come from RANS simulations whereas the quantities $c^\NN_k$, $k = 1,\dots,N_c$, are the Neural Network outputs. In equation \eqref{eq:to_minimize}, with an abuse of notation, we define $(\widetilde{\divtau})^\DNS=\frac{(k^{1/2})^\RANS}{\varepsilon^\RANS}\divtau^\DNS$. In most cases the DNS $k$ and $\varepsilon$ fields are not available. For this reason, the dimensionless VBNN output must be successively dimensionalized using RANS fields. 

The quantity $(\divtau)^\DNS$ is obtained by interpolation of the available $\btau^\DNS$ into the RANS mesh followed by computation of its divergence on the RANS mesh. 

\subsubsection{Architecture and hyperparameters} 
\begin{figure}[]
	\centering
	\includegraphics[width=1.0\textwidth]{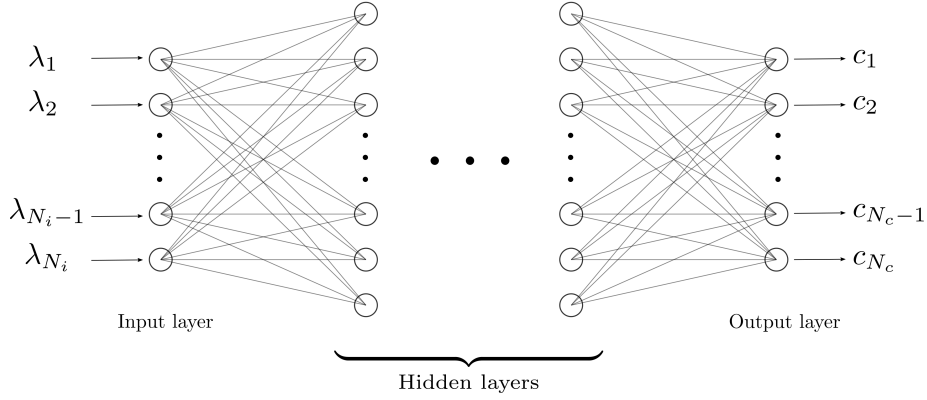}
	\caption{Architecture of the Vector Basis Neural Network.}
	\label{Fig:VBNN_architecture}
\end{figure}

The input and output layers of the VBNN have a number of nodes that is constrained by the assumption \eqref{eq:divtau_const_hp}. In particular, the input layer has $N_i$ nodes (as many as the invariants) while the output layer has $N_c$ nodes (as many as the coefficients to be predicted), see Figure \ref{Fig:VBNN_architecture}. After some tests, it has been noted that the network accuracy is not particularly sensible to both network depth and width. This behaviour was observed also in \cite{Milani2020}. At the end, 8 hidden layers have been defined with 30 nodes each as made in \cite{Ling2016}.

It has been observed that the network is not affected by overfitting issues. Thus, the regularization term associated to the weights norm has been shut down. The Adam optimizer \cite{Adam} is used with learning rate that decreases during the training stage from $10^{-3}$ to $10^{-5}$ and batch size equal to 50. The Exponential Linear Unit (ELU) \cite{ELU} function has been chosen as activation function because of the better observed performances. 

Due to the intrinsic stochasticity of the optimization process, several training runs with the same hyperparameters have been performed. Among them, the run that minimised the validation error has been identified as the reference one for the specific hyperparameters choice.

\subsection{Invariance properties}
\subsubsection{Galilean invariance}
All the inputs of the VBNN are Galilean invariant. Consequently, the outputs of the VBNN, that depends on the inputs, do not change through a Galilean transformation.

\subsubsection{Coordinates-frame rotation invariance}
It is well known that the representations of scalars $s$, vectors $\mathbf{v}$ and second-order tensors $\mathbf{T}$ follow the transformation laws
\begin{equation} \label{rotation}
s^{\mathbf{Q}} = s, \qquad \mathbf{v}^{\mathbf{Q}} = \mathbf{Q}\mathbf{v}, \qquad \mathbf{T}^{\mathbf{Q}} = \mathbf{Q}\mathbf{T}\mathbf{Q}^T,
\end{equation}
for any rotation matrix $\mathbf{Q}$. The apex $\mathbf{Q}$ denotes the representation of the quantity in the rotated coordinates system.

The VBNN is coordinates-frame rotation invariant in the sense that all the scalar outputs are coordinates-frame rotation invariant. This property directly derives from the invariance of the scalar inputs, i.e. $\lambda^{\mathbf{Q}}_i = \lambda_i$. It implies that $\widetilde{\divtau}$ transforms correctly under rotations. Indeed
\begin{equation} \label{eq:divtau_rotation}
\begin{split}
\widetilde{\divtau}^{\mathbf{Q}}  = &
\sum_{k=1}^{N_c}{c_k (\lambda^{\mathbf{Q}} _1,\dots,\lambda^{\mathbf{Q}} _{N_i}) \ \mathbf{t}^{\mathbf{Q}} _k}  = 
\sum_{k=1}^{N_c}{c_k (\lambda_1,\dots,\lambda_{N_i}) \ \mathbf{Q} \mathbf{t}_k} = \\
= &\mathbf{Q} \Big[ \sum_{k=1}^{N_c}{c_k (\lambda_1,\dots,\lambda_{N_i}) \ \mathbf{t}_k} \Big]
= \mathbf{Q} \widetilde{\divtau}.
\end{split}
\end{equation}

\subsection{Implicit-Explicit treatment of $\divtau$}
Once the term $\divtau$ is obtained, the RANS system \eqref{eq:RANS} has to be solved. The easiest approach is to treat explicitly this term like a source term. However in \cite{Wu2019}, in the data-driven Reynolds Stress Tensor setting, the ill-conditioning of this approach is highlighted. In the former work, the authors propose to treat implicitly the Reynolds Stress Tensor component aligned to the mean strain rate tensor $\mathbf{S}$ into the diffusive term. A better conditioning of the system was observed with this approach. The present work takes inspiration on this remark with the slight change imposed by dealing with the divergence of the RST instead of the RST itself. Hence, the attention will be devoted to the term aligned with $\nabla \cdot \mathbf{S}$.

Let take the expression \eqref{eq:divtau_expansion} with the first term explicitly written
\begin{equation}
\widetilde{\divtau} = c_1 \divs +  \sum_{k=2}^{N_c}{c_k  \ \mathbf{t}_k}.
\end{equation}
Recalling $\widetilde{\divtau} = \frac{k^{1/2}}{\varepsilon}\divtau$ and $\widetilde{\nabla \cdot \mathbf{S}} = \frac{k^{5/2}}{\varepsilon^2} \nabla \cdot \mathbf{S}$, the above expression becomes
\begin{equation} \label{eq:divtau_with_first_term}
\divtau = \frac{k^2}{\varepsilon} \ c_1 \nabla \cdot \mathbf{S} + \frac{\varepsilon}{k^{1/2}} \sum_{k=2}^{N_c}{c_k  \ \mathbf{t}_k}.
\end{equation}
The scalar term $\frac{k^2}{\varepsilon}$ is dimensionally a viscosity. This remark drives to the definition of the \emph{turbulent-like viscosity}
\begin{equation} \label{eq:nut_def}
\nu_{tl} := - \frac{k^2}{2 \varepsilon} \ c_1.
\end{equation}
Thus, the momentum equation of the RANS system reads
\begin{equation} \label{eq:RANS_momentum_nutl_not_splitted}
\frac{\partial \vel}{\partial t}+\vel \cdot \nabla\vel - (\nu+\nu_{tl}) \Delta \vel = - \nabla p - \frac{\varepsilon}{k^{1/2}} \sum_{k=2}^{N_c}{c_k  \ \mathbf{t}_k}.
\end{equation}
Looking to the obtained system, the difference between the turbulent-like and the turbulent viscosity consists in their positioning with respect to the divergence operator. Indeed the former is located outside the divergence, i.e. $\nu_{tl} \nabla \cdot (\nabla \vel)$, while the former inside it, i.e. $\nabla \cdot (\nu_t \nabla \vel)$, see \eqref{eq:RANS} and \eqref{eq:linear_closures}.

In general, it is not guaranteed that $\nu_{tl}>0$ (corresponding to $c_1<0$) holds everywhere. Let us write $\nu_{tl} = \nu^+_{tl} + \nu^-_{tl}$ where $\nu^+_{tl}(x) = \max(\nu_{tl},0)$ is the positive part of the turbulent-like viscosity. Finally, let define 
\begin{equation} \label{eq:divtau_perp}
(\divtau)^{\dag} := - \nu^-_{tl} \nabla \cdot \mathbf{S} + \frac{\varepsilon}{k^{1/2}} \sum_{k=2}^{N_c}{c_k  \ \mathbf{t}_k} .
\end{equation}
The final RANS system with Implicit-Explicit treatment reads
\begin{equation} \label{eq:Implicit_Explicit_RANS}
\begin{dcases}
&\nabla \cdot \vel=0, \\
&\frac{\partial \vel}{\partial t}+\vel \cdot \nabla\vel - (\nu+\nu^+_{tl}) \Delta \vel = - \nabla p - (\divtau)^{\dag},
\end{dcases}
\end{equation} 
where the term associated to $\nu^+_{tl}$ is treated implicitly into the diffusion term while the term $(\divtau)^{\dag}$ is treated explicitly.

More details about the implementation of the Implicit-Explicit treatment in \openfoam are given in Appendix.

%
%

\section{Numerical results} \label{Sec:Results}
This section discusses about the application of the VBNN into two classical benchmark flows: the flow in a square duct and the flow over periodic hills. As it will be discussed, standard RANS models fails in the description of the velocity field in these configurations.

The VBNN is implemented and trained in Python using the Tensorflow package \cite{tensorflow2015-whitepaper} while all the RANS computations are performed with the Finite Volume Method based \openfoam opensource code \cite{OpenFOAM}.

%
%

\subsection{Flow in a square duct}

\subsubsection{Dataset} 
\begin{figure}[]
	\centering
	\includegraphics[width=0.8\textwidth]{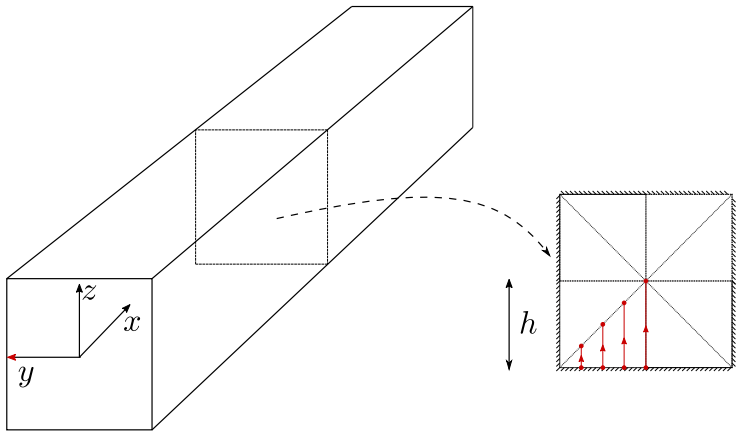} 
	\caption{Square duct domain.}
	\label{Fig:domain_section}
\end{figure}

In \cite{Pinelli2010} DNS data are provided at several bulk Reynolds numbers $Re_b$. The simulations with $Re_b = 2200, 2600, 2900$ are employed for training purposes. In particular, $80\%$ of the data are used for training while the remaining $20\%$ for validation. The flow at $Re_b = 3500$ is employed to test the network prediction ability. The test flow Reynolds number is higher than the training ones to analyse the extrapolation property of the VBNN. This particular flow is (in average) stationary and uniform across the main streamwise direction. Only the data coming from three square sections in the central region of the duct are used to reduce considerably the training effort. Figure \ref{Fig:domain_section} shows the domain and one square section. The obtained dataset counts roughly $2 \cdot 10^4$ simulation cells. The Reynolds Stress Transport Model \cite{LRR} is used as RANS model. In the following it will referred as Baseline.

\subsubsection{Results analysis}
\begin{figure}[]
	\centering
	\subfigure[$(\widetilde{\divtau})_1$] 
{\includegraphics[width=1.0\textwidth] {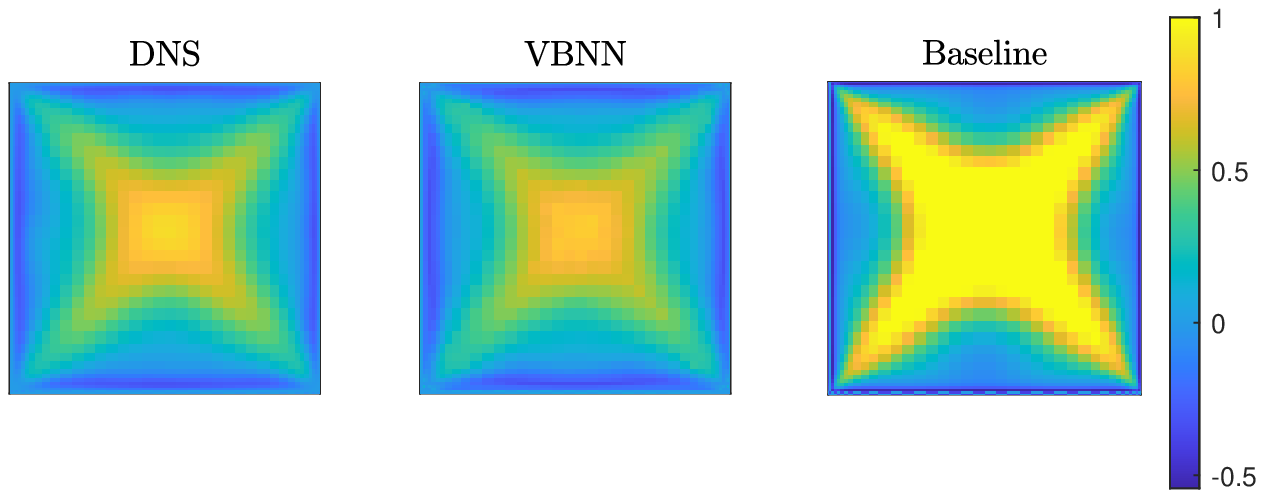}}
\subfigure[$(\widetilde{\divtau})_2$]
{\includegraphics[width=1.0\textwidth] {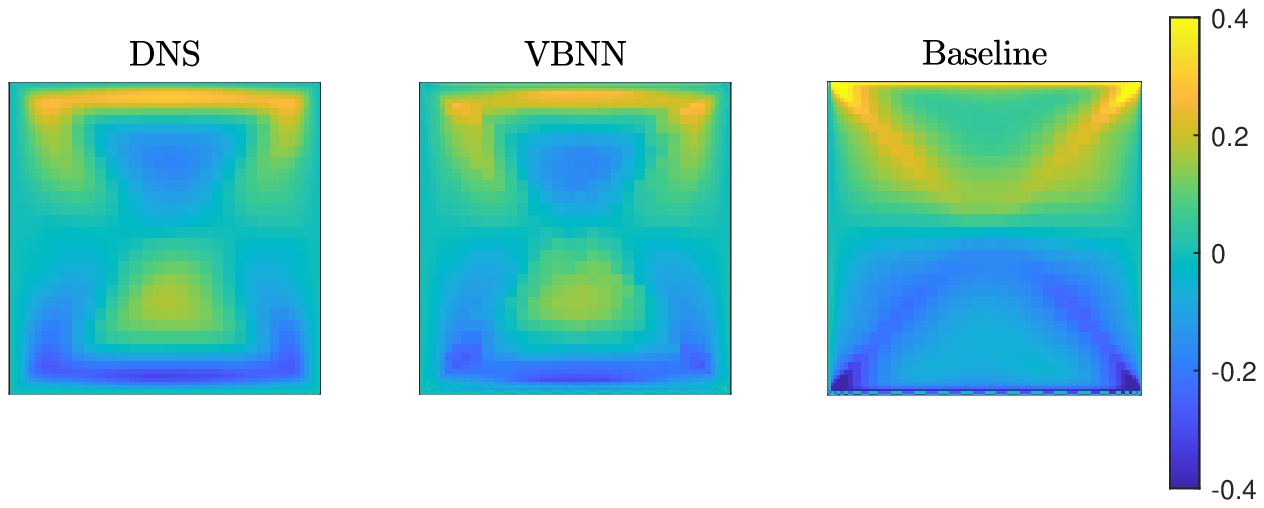}}
\subfigure[$(\widetilde{\divtau})_3$]
{\includegraphics[width=1.0\textwidth] {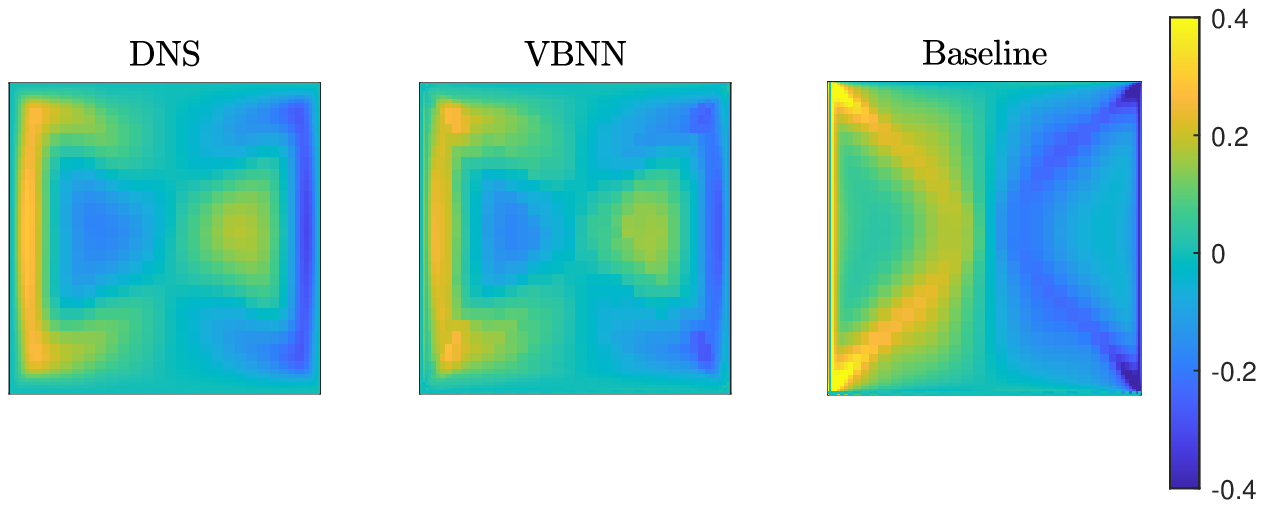}}
\caption{Comparison between the components of $\widetilde{\divtau}$ from DNS (on the left), VBNN (in the middle) and RSTM Baseline one (on the right).}
\label{Fig:divtau_divs_gradk_no_normu}
\end{figure}

Figure \ref{Fig:divtau_divs_gradk_no_normu} compares the components of the vector $\widetilde{\divtau}$ obtained from DNS, VBNN and Baseline model respectively. Regarding the first component, the VBNN is in agreement with the DNS both qualitatively and quantitatively while the Baseline overpredicts it in the center and along the diagonals of the square section. Regarding the second and third components, the Baseline have positive and negative values located in two separated square section's halves. In addition maxima and minima are overestimated in absolute value. On the other hand, the VBNN describes correctly the values of these components and where these are positive and negative. However, VBNN predicts in few cells near the corners maxima or minima that are not in the DNS.  

Table \ref{Tab:RMSE_gradk_vs_baseline} shows the Root Mean Square Error (RMSE) defined as 
\begin{equation}
\textnormal{RMSE} = \sqrt{\frac{1}{3 N_{cells}} \sum_{i=1}^{N_{cells}} {\vert \vert \widetilde{\divtau}_i^\DNS - \widetilde{\divtau}_i^{\textnormal{model}} \vert \vert^2}},
\end{equation}
where $N_{cells}$ is the number of cells in the RANS square section grid. This metric quantitatively measures the distance between the DNS dimensionless target and the turbulence model ones. The Baseline RMSE is one order of magnitude higher than the VBNN.

\begin{table}[]
\caption{Root Mean Square Error (RMSE) of VBNN and Baseline models using $\widetilde{\divtau}^\DNS$ as reference.}
\begin{center}
{\renewcommand{\arraystretch}{1} 
\begin{tabular}{*{2}{c}}
\hline
model & RMSE \\
\hline
VBNN & 0.32 e-1\\
Baseline & 2.43 e-1 \\
\hline
\end{tabular}}
 \label{Tab:RMSE_gradk_vs_baseline}
\end{center}
\end{table}

\begin{figure}[]
	\centering
	\includegraphics[width=0.4\textwidth]{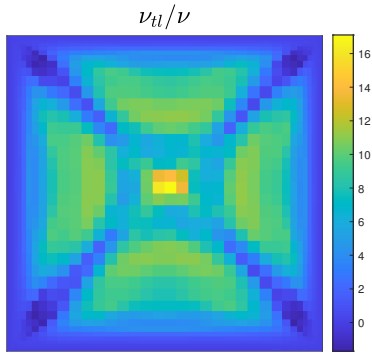}
	\caption{Ratio between turbulent-like viscosity $\nu_{tl}$ and kinematic viscosity $\nu$.}
	\label{Fig:nutl_divs_gradk_no_normu_adim}
\end{figure}

Among all the predicted coefficients in the expansion \eqref{eq:divtau_expansion}, the first one plays a key role in the conditioning of the RANS system. In particular, the more extended are the regions with a negative predicted first coefficient (and consequently positive $\nu_{tl}$) and the bigger in magnitude are these negative values, the better conditioned is the system. Figure \ref{Fig:nutl_divs_gradk_no_normu_adim} shows the ratio between the turbulent-like viscosity $\nu_{tl}$, defined in \eqref{eq:nut_def}, and the kinematic viscosity $\nu$. The ratio is positive in the majority of the square section with values bigger than 8 frequently occurring. The negative regions are very limited and located on the square diagonals near the corners. The minimum value of the ratio is lower that $-1$, in particular $\min(\nu_{tl}/\nu) = -1.70$. If the turbulent-like viscosity was treated completely implicitly, the total viscosity associated to the laplacian operator in \eqref{eq:RANS_momentum_nutl_not_splitted} would be negative in some regions. This observation justifies the splitting of $\nu_{tl}$ into its positive and negative part being the former only treated implicitly.

\begin{figure}[]
	\centering
	\includegraphics[width=1.0\textwidth]{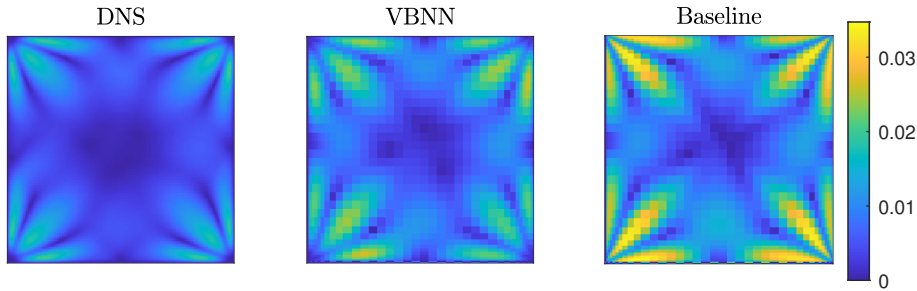}
	\caption{Magnitude of the secondary motion in the DNS (left), the VBNN (middle) and the Baseline model (right). Lighter colors correspond to higher magnitudes of the secondary flow.}
	\label{Fig:secondary_motion}
\end{figure}

The obtained data-driven $\divtau$ is successively inserted into the RANS solver to obtain new steady fields. Figure \ref{Fig:secondary_motion} shows the magnitude of the secondary motion $\vert \vert(u_y,u_z)^T\vert \vert_2/u_b$ (assuming the streamwise velocity aligned to the $x$ axis), where $u_b$ is the bulk velocity. Lighter colors correspond to higher values of the norm. The different resolution between the models is due to the grid density, being the DNS one much finer than the VBNN and Baseline one (the same grid is employed for both models). Even if the Baseline model describes correctly the regions where the secondary motion is more prominent, it drastically overpredicts it. On the other hand, the VBNN secondary motion is still overpredicted, but its magnitude is in between the DNS case and the Baseline one. To make a quantitative comparison, Table \ref{Tab:second_motion} reports the values of $\max(\vert \vert(u_y,u_z)^T\vert \vert_2)/u_b$, and the relative amplification using the DNS value as reference. The VBNN approach reduces the overestimation from $70\%$ of the Baseline model to $30\%$. Finally, the VBNN secondary motion is characterized by symmetry (up to numerical discrepancies in the central and on the peaks regions) with respect to square section diagonals, while the Baseline case does not correctly respect the symmetry.

\begin{table}[]
\caption{Maxima of the secondary motion norm and corresponding amplification factor}
\begin{center}
\begin{tabular}{*{3}{c}} 
\hline
model & $\max(\vert \vert(u_y,u_z)^T\vert \vert_2)/u_b$ & $\frac{\max(\vert \vert(u_y,u_z)^T\vert \vert_2)}{\max(\vert \vert(u_y^\DNS,u_z^\DNS)^T\vert \vert_2)}$ \\
\hline
DNS & 2.04 e-2 & 1 \\
VBNN & 2.65 e-2 & 1.30 \\
Baseline & 3.47 e-2 & 1.70 \\
\hline
\end{tabular}
\label{Tab:second_motion}
\end{center}
\end{table}

\begin{figure}[]
	\centering
	\subfigure[] 
{\includegraphics[width=0.65\textwidth] {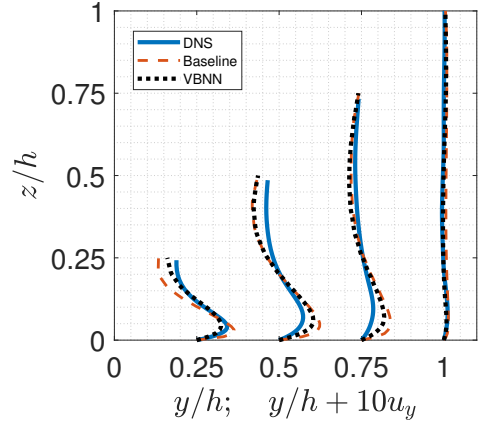}}
\subfigure[]
{\includegraphics[width=0.65\textwidth] {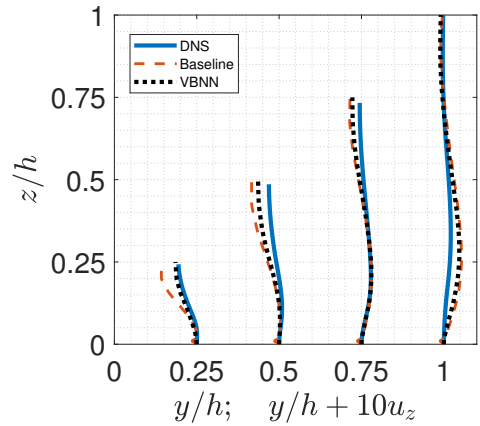}}
\caption{Secondary motion velocity components along the red lines defined in the square section in Figure \ref{Fig:domain_section}. The sign of $u_y$ is changed compared to the coordinates defined in Figure \ref{Fig:domain_section} to make comparison with \cite{Wu2018,Wu2019} easier.}
\label{Fig:velocity_profiles}
\end{figure}

Figure \ref{Fig:velocity_profiles} shows the $u_y$ and $u_z$ profiles along the red lines defined in Figure \ref{Fig:domain_section} in the lower-left square section quadrant. The $u_y$ has been flipped of sign to make comparison with \cite{Wu2018,Wu2019} easier. The Baseline model overpredicts the magnitudes of both velocity components. On the other hand, the VBNN curves close to the corners, i.e. for $y/h = 0.25$, almost overlaps the DNS ones. Small improvements are noticeable also for the other curves, in particular near wall for $u_y$ curves and far from wall for the $u_z$ ones.

\subsubsection{Role of the Implicit-Explicit treatment}
\begin{figure}[]
	\centering
	\includegraphics[width=0.6\textwidth]{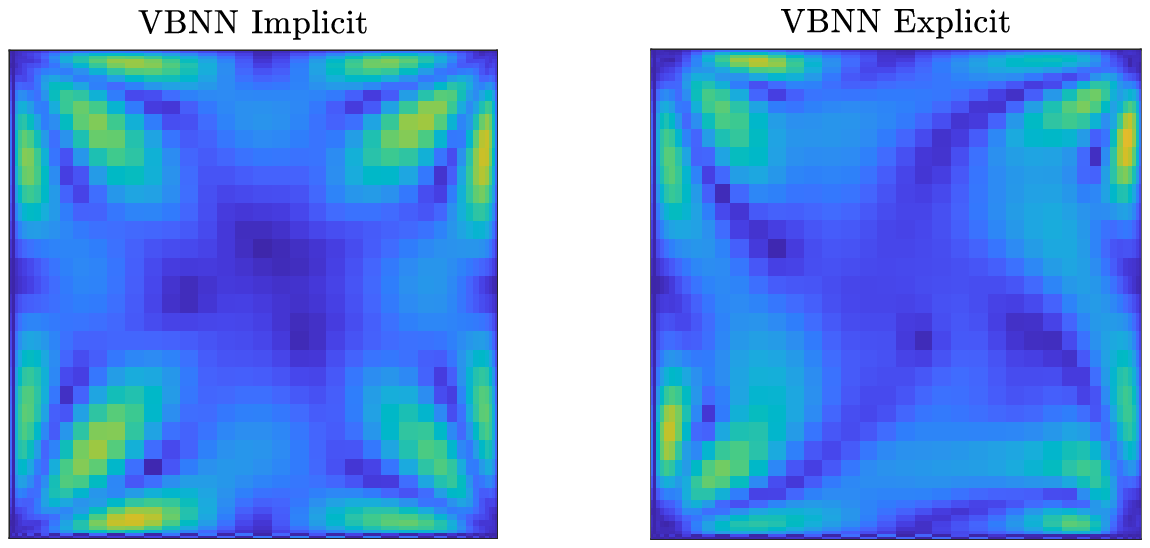}
	\caption{Comparison of the magnitude of the secondary motion in case of Implicit-Explicit (left) or purely Explicit (right) treatment of $\divtau$. The same colormap as in Figure \ref{Fig:secondary_motion} is used.}
	\label{Fig:secondary_motion_gradk_Implicit_vs_Explicit}
\end{figure}

Figure \ref{Fig:secondary_motion_gradk_Implicit_vs_Explicit} compares the secondary motion obtained with the Implicit-Explicit treatment of the divergence of the RST and with the totally Explicit one. Even if the latter damps the magnitude of the motion as desired, it unphysically breaks the symmetries. In addition, the Explicit simulation takes an order of magnitude more time steps to reach the steady state. In general, a convergence speed up of implicit treatments was reported in \cite{Thompson2021}. 

It is important to highlight that differences between the two fields are uniquely due to the treatment of the divergence of the RST into the equations. As a matter of fact, the same $\divtau$ field is fed into the equations. 

\subsubsection{Role of the dependencies choice}
In this section we test also the dependences hypothesis \eqref{eq:divtau_const_hp} by choosing the simpler relation 
\begin{equation}
\widetilde{\divtau} = \mathbf{f}(\mathbf{s},\mathbf{w},\widetilde{\nabla \cdot \mathbf{S}}, Re_d),
\end{equation}
i.e. by removing the dependence on $\gradk$. In this case the basis consists of 6 elements while the invariants are 14. This hypothesis still let possible the Implicit-Explicit treatment of the RANS system because $\divs$ is still a basis vector.

The RMSE error in this case is $0.37 \cdot 10^{-1}$, bigger that the corresponding value in Table \ref{Tab:RMSE_gradk_vs_baseline}. This behaviour is expected because a smaller vector basis and a smaller set of invariants are considered.

\begin{figure}[]
	\centering
	\includegraphics[width=0.6\textwidth]{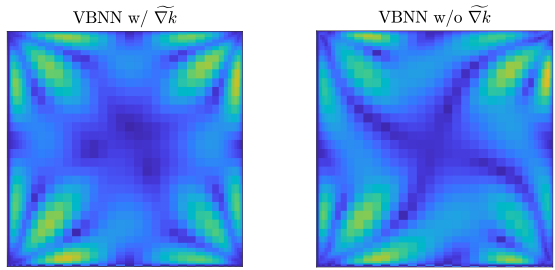}
	\caption{Magnitude of the secondary motion: on the left the field assuming a dependence on $\widetilde{\nabla k}$, on the right the field  not assuming a dependence on $\widetilde{\nabla k}$. The same colormap as in Figure \ref{Fig:secondary_motion} is used.}
	\label{Fig:secondary_motion_with_gradk_vs_without_gradk}
\end{figure}

Figure \ref{Fig:secondary_motion_with_gradk_vs_without_gradk} shows the secondary motion in the two cases. Even when $\gradk$ is not considered, the secondary motion is still correctly damped compared to the Baseline case. However, the field loses its symmetry with respect to the square diagonals. This fact can be ascribed to the worse description of the explicit term $(\divtau)^{\dag}$ in the RANS system.

%
%

\subsection{Flow over periodic hills}

\subsubsection{Dataset}
\begin{figure}[]
	\centering
	\includegraphics[width=0.7\textwidth]{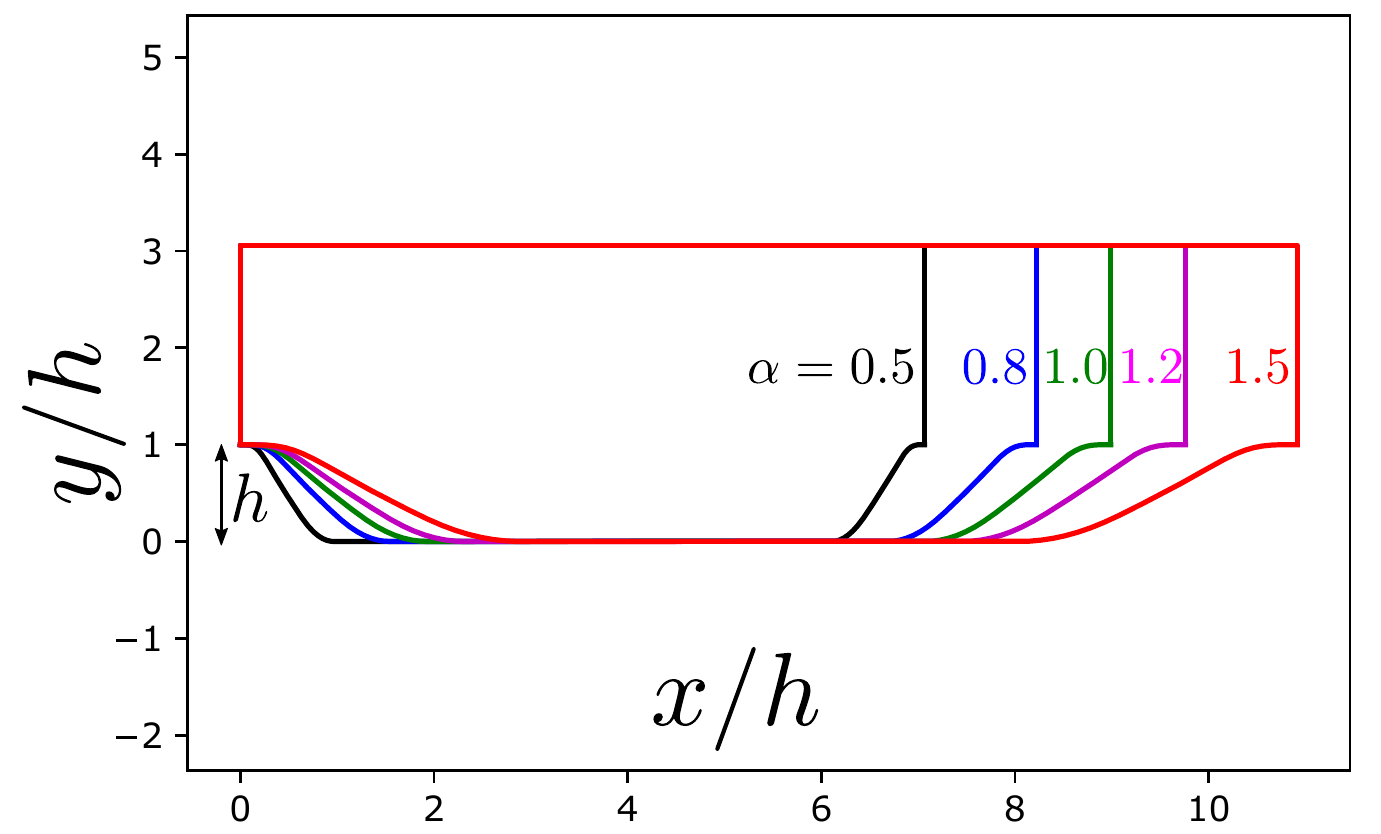}
	\caption{Periodic hills shapes with respect to the $\alpha$ slope parameter.}
	\label{Fig:hills_shape}
\end{figure}

The DNS data coming from \cite{Xiao2020} have been used where several simulations with different hills geometries but fixed bulk Reynolds number are available. Figure \ref{Fig:hills_shape} shows the different steepness associated to the parameter $\alpha$. The smaller is $\alpha$, the steeper is the hill profile. All lengths are set dimensionless dividing by the hill high $h$.

The flows with $\alpha = 0.5, 0.8, 1.2, 1.5$ have been employed for the training (with the same splitting $80\% - 20\%$ for training and validation) while the case with $\alpha = 1.0$ is used for testing only. The RANS simulations have been performed with a 2D domain. The cardinality of the training dataset is about $5.9 \cdot 10^4$ cells that is the cells number in the 2D RANS domain. It has been observed that a dropout regularisation method \cite{dropout} with drop probability of 0.1 was helpful in the training and consequently adopted. The Launder and Sharma $k-\varepsilon$ linear model \cite{Launder1974} is employed as Baseline RANS model. The \openfoam case was already available in \cite{Xiao2020}.

\subsubsection{Results analysis}

\begin{figure}[]
	\centering
	\subfigure[$(\widetilde{\divtau})_1$] 
{\includegraphics[width=1.0\textwidth] {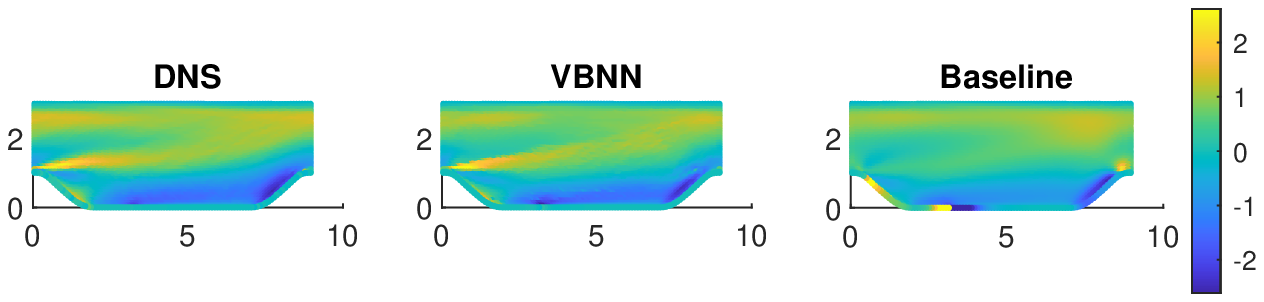}} 
\subfigure[$(\widetilde{\divtau})_2$]
{\includegraphics[width=1.0\textwidth] {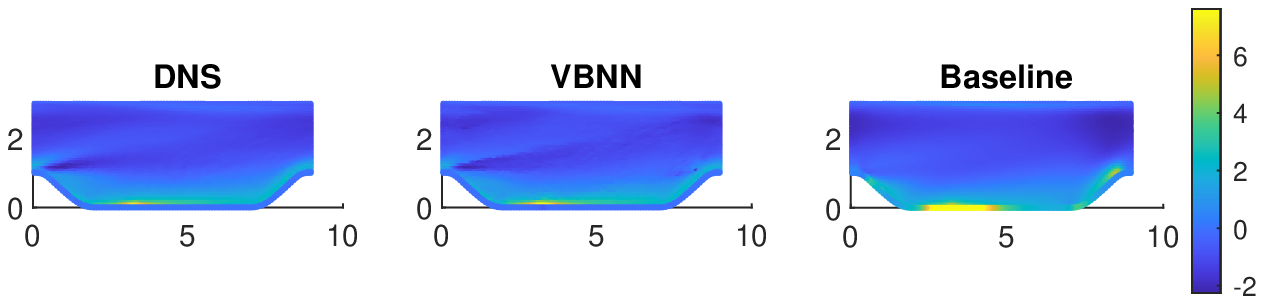}} 
\caption{Comparison between the components of $\widetilde{\divtau}$ from the DNS (on the left), the components obtained with the VBNN (in the middle) and the RSTM Baseline one (on the right). The third component is not shown because uniformly zero.}
\label{Fig:divtau_periodic_hills}
\end{figure}

Figure \ref{Fig:divtau_periodic_hills} shows the first two components only of $\widetilde{\divtau}$ for the DNS, VBNN and Baseline models, being the third component zero (the VBNN correctly predicts it). The VBNN components are in agreement with the reference ones while this is not true for the Baseline model. 

For the first component, VBNN provides a correct description of the maxima loci that starts from the crest of the front hill. It also predicts the limited region of local maximum at the middle-end of the first hill and the following minimum. The Baseline case has wrong, both in location and values, maxima and minima in the left part of the domain. It predicts correctly the minima region on the second hill (well predicted by the VBNN model too). However, there is a wrong maximum on the top of the second hill. 

Regarding the second component, the VBNN model describes correctly the value and the extension of the maximum located at $x/h \approx 3.5, \ y/h = 0$ and the Baseline model overpredicts both aspects. Finally the VBNN case is characterized by the minima region that starts from  the first hill crest.

\begin{figure}[]
	\centering
	\includegraphics[width=0.6\textwidth]{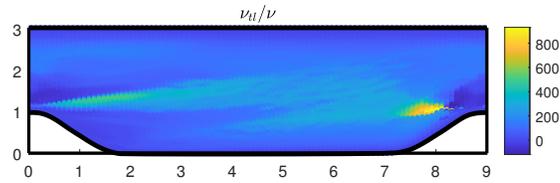} 
	\caption{Ratio between turbulent-like viscosity $\nu_{tl}$ and kinematic viscosity $\nu$.}
	\label{Fig:nutl_periodic_hills}
\end{figure}

Figure \ref{Fig:nutl_periodic_hills} shows $\nu_{tl}/ \nu$ to understand the relevance of the Implicit-Explicit treatment. The ratio is positive in the majority of the domain with the exception of the two region immediately above the hills, in particular above the rear one. In particular, this quantity assumes values $O(10^2)$ with maximum of about 800. This behaviour is helpful for the conditioning of the system. It has been observed that the dropout regularisation helps in reducing the regions with negative turbulent-like viscosity.

\begin{figure}[]
	\centering
	\subfigure[] 
{\includegraphics[width=0.8\textwidth] {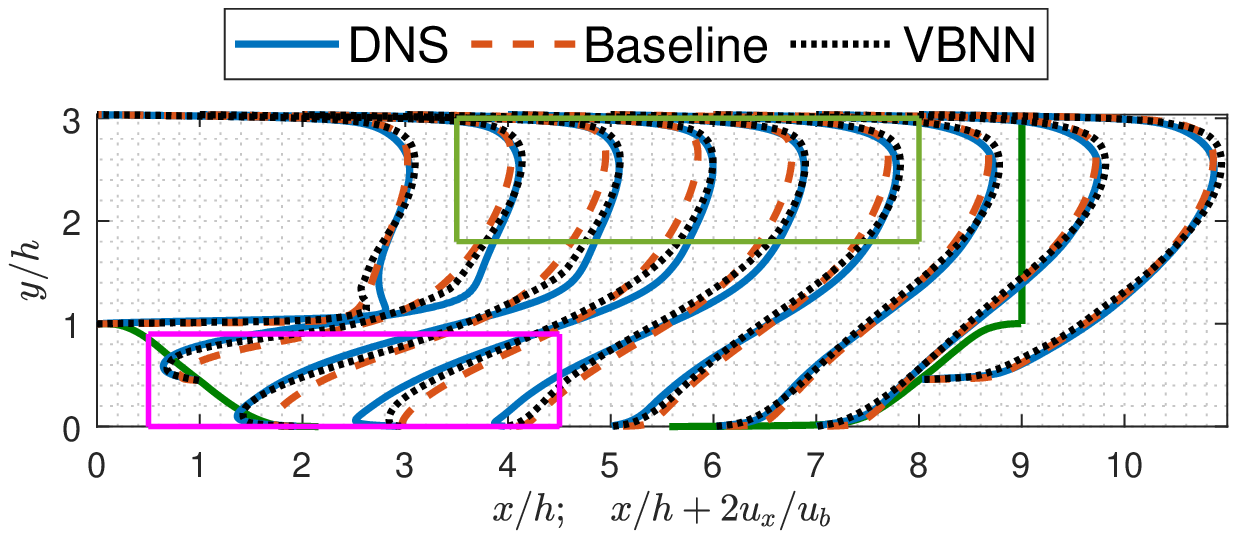}} 
\subfigure[]
{\includegraphics[width=0.4\textwidth] {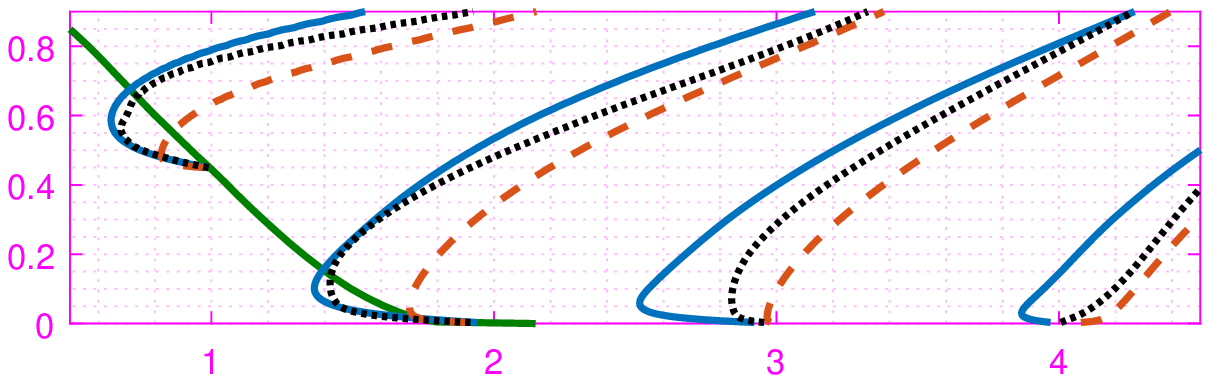}} 
\subfigure[]
{\includegraphics[width=0.4\textwidth] {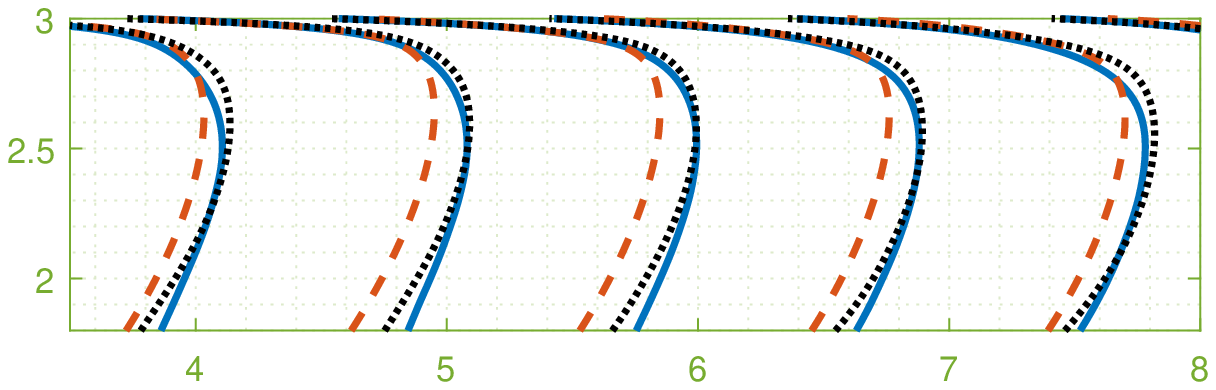}}
\caption{$u_x$ profiles comparison between DNS, VBNN and Baseline cases (a). Zoom downstream the first hill crest in magenta box in figure above (b). Zoom in the middle top wall region in green box in figure above (c).}
\label{Fig:hills_ux_profiles_with_zoom}
\end{figure}

Figure \ref{Fig:hills_ux_profiles_with_zoom} represents the horizontal velocity profiles along the vertical lines at $x/h = c$ with $c = 0,\dots, 8$. The profiles are obtained once the simulation reaches the steady state with $\divtau$ coming from the VBNN model. Figure \ref{Fig:hills_ux_profiles_with_zoom}(a) shows the whole domain while Figure \ref{Fig:hills_ux_profiles_with_zoom}(b) focuses on the first hill downstream wall region and Figure \ref{Fig:hills_ux_profiles_with_zoom}(c) depicts the middle top wall region. 

Generally speaking, the VBNN curves are closer to the DNS ones compared to the Baseline ones. This behaviour is observable in the whole computational domain. It is worth mentioning that the VBNN model predicts, even if underestimated, the local maximum of the horizontal velocity on the crest of the first hill ($x/h = 0$, $y/h=1$). This behaviour is not captured by the Baseline model for which $u_x$ monotonically increases until $y/h \approx 2.8$. 

Figure \ref{Fig:hills_ux_profiles_with_zoom}(b) shows that the Baseline model predicts almost null reversal flow downstream the first hill. At the contrary, the VBNN simulation predicts reversal flow quantitatively close to the DNS case for $x/h = 1$ and $x/h = 2$. VBNN still predicts reversal flow, even if underestimated, for $x/h = 3$. However, at $x/h = 4$ the VBNN horizontal velocity is positive near wall while the DNS one is still negative.  

Figure \ref{Fig:hills_ux_profiles_with_zoom}(c) represents the maxima of the curves that are located near the upper wall ($y/h = 3.036$). The maxima are slightly but constantly underestimated by the Baseline model. Conversely, the VBNN curves are very close to the DNS ones and do not suffer of the underestimation issue.

\begin{figure}[]
	\centering
	\subfigure[] 
{\includegraphics[width=0.8\textwidth] {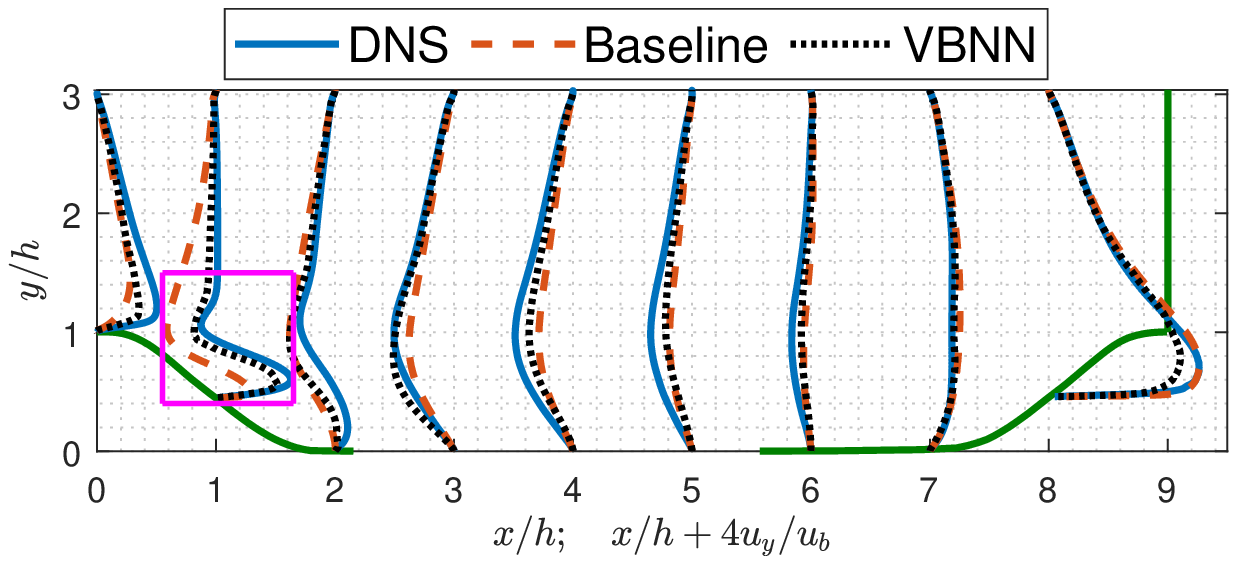}} 
\subfigure[]
{\includegraphics[width=0.4\textwidth] {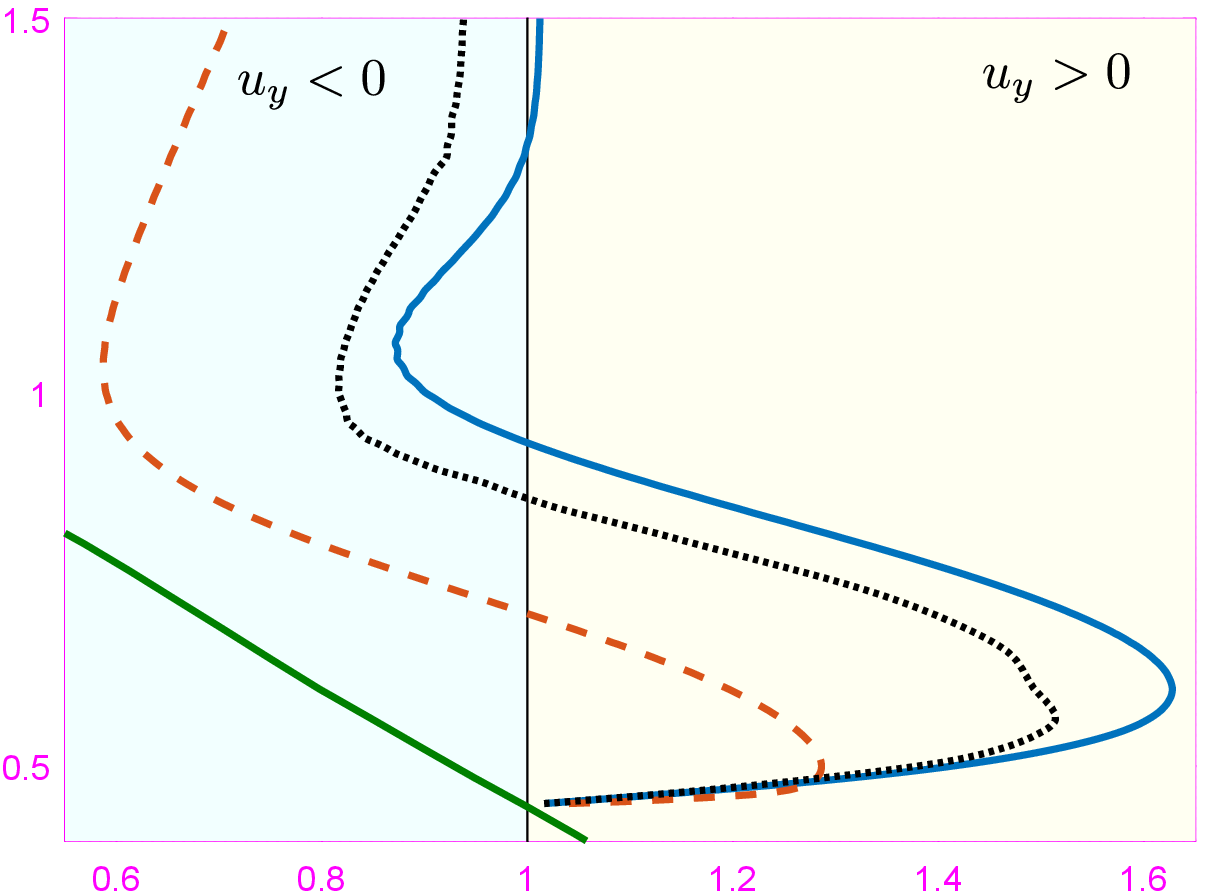}} 
\caption{$u_y$ profiles comparison between DNS, VBNN and Baseline cases (a). Zoom downstream the first hill crest at $x/h = 1$ in magenta box in figure above (b).}
\label{Fig:hills_uy_profiles_with_zoom}
\end{figure}

Figure \ref{Fig:hills_uy_profiles_with_zoom}(a) represents the vertical velocity profiles along the vertical lines at $x/h = c$ with $c = 0,\dots, 8$, while Figure \ref{Fig:hills_uy_profiles_with_zoom}(b) focuses on the $x/h=1$ line near the hill. VBNN leads to improvements compared to the Baseline $k-\varepsilon$ model, even if less prominent than the $u_x$ case. The region with best improved accuracy is the $x/h = 1$ line near the hill wall represented in  Figure \ref{Fig:hills_uy_profiles_with_zoom}(b). The VBNN model predicts a positive velocity region close to the DNS one while the $k-\varepsilon$ turbulence model predicts a positive velocity in a smaller region. In particular, the vertical velocity goes from negative to positive at $y/h \approx 0.93$ in the DNS, $y/h \approx 0.86$ in the VBNN and $y/h \approx 0.70$ in the $k-\varepsilon$ model.

%
%

\section{Conclusions} \label{Sec:Conclusions}
The present paper proposes a new data-driven turbulence model to close and increase accuracy of the RANS equations. Hence the model predicts the divergence of the Reynolds Stress Tensor, called Reynolds Force Vector in \cite{Thompson2019}.
This target vectorial quantity is obtained through a Neural Network that predicts the coefficients of a vector basis expansion. These coefficients are functions of invariant scalar quantities that depend on the averaged fields. The vector basis and the invariants are uniquely defined once a list of dependencies of the target function is made. A possible dependencies choice is discussed starting from a well known assumption in literature for the anisotropic Reynolds Stress Tensor.
Moreover, the architecture of the trained network and the invariants choice guarantee both Galilean and coordinates-frame rotation invariances. In addition, this approach closes directly the RANS system and does not require any coupling with classic turbulence models. An implicit treatment of the first term of the expansion is proposed to increase the conditioning of the RANS system.

The proposed model is tested for the flow in a square duct and the flow over periodic hills. Both flows, despite their geometrical simplicity, present features that classic turbulence models do not describe correctly, in particular the secondary flow for the former and the recirculation flow downstream the first hill in the latter. The data-driven model qualitatively and quantitatively outperforms classic turbulence models in both scenarios.

%
%

\section*{Acknowledgements}
D. Oberto wants to thank also M. Pintore and F. Della Santa for the precious advices on the neural networks implementation and A. Giammarini for the discussion on invariance properties. The authors are members of the Italian INdAM-GNCS research group.

%
%

\section*{Appendix} \label{Sec:Appendix}

\subsection*{Dependencies of $\widetilde{\divtau}$} 
Starting from the definition of anisotropic RST and its expansion in \eqref{eq:aRST_expansion}, one could write
\begin{equation} \label{eq:tau_starting_point}
\btau = 2 k \Big[\sum_{j=1}^{10} {c_j (\lambda_1,\dots,\lambda_6) \ \mathbf{T}_j} + \frac{1}{3} \mathbf{I}\Big].
\end{equation}
It is possible to compute the divergence of the above expression as
\begin{equation} \label{eq:tau_after_}
\begin{split}
\divtau = 
&2 \underbrace{\sum_{j=1}^{10} {c_j(\lambda_1,\dots,\lambda_6) [\nabla k]^T \mathbf{T}_j}}_{\alpha} +
2 \underbrace{\sum_{j=1}^{10} {k [\nabla c_j(\lambda_1,\dots,\lambda_6)]^T  \mathbf{T}_j}}_{\beta} + \\
& 2 \underbrace{\sum_{j=1}^{10} {k c_j(\lambda_1,\dots,\lambda_6) \nabla \cdot \mathbf{T}_j}}_{\gamma} +
\underbrace{\frac{2}{3} \nabla k}_{\delta}.
\end{split}
\end{equation}

Let suppose that both the tensor basis expressed in \eqref{eq:basis_a} and the turbulent kinetic energy are known (the standard approach is to obtain them from a RANS simulation). While the terms $\alpha, \gamma, \delta$ can be treated by a neural network that obtains scalar coefficients only, the term $\beta$ contains the gradient of the unknowns that are vector quantities. Thus, the expression \eqref{eq:tau_after_} can not be directly used while preserving coordinates-frame rotation invariance. Nonetheless, it can be used as a guideline to write down a new constitutive law for $\divtau$ or its dimensionless counterpart  $\widetilde{\divtau} = \frac{k^{1/2}}{\varepsilon} \divtau$. It has been decided to predict the latter to be as close as possible to the TBNN approach in \cite{Ling2016} where the dimensionless anisotropic Reynolds Stress Tensor is predicted.

From the above computations, it seems natural to suppose dependences of $\widetilde{\divtau}$ from $\mathbf{s}$, $\mathbf{w}$, their respective divergences and $\nabla k$. It is worth noting that the divergence of each tensors $\{\mathbf{T}_j\}$ in \eqref{eq:tau_after_} involves multiplications of $\mathbf{s}$, $\mathbf{w}$ and their respective divergences. 

In order to work with dimensionless quantities only, analogously to \cite{Ling2016}, it has been decided to make the following assumption:

\begin{equation} \label{eq:divtau_const_hp_not_final}
\widetilde{\divtau} = \mathbf{f}(\mathbf{s},\mathbf{w},\widetilde{\nabla \cdot \mathbf{S}},\widetilde{\nabla \cdot \mathbf{W}},\widetilde{\nabla k}),
\end{equation}
where $\mathbf{s} = \frac{k}{\varepsilon}  \mathbf{S}$, $\mathbf{w} = \frac{k}{\varepsilon}  \mathbf{W}$, $\widetilde{\nabla \cdot \mathbf{S}} = \frac{k^{5/2}}{\varepsilon^2}  \nabla \cdot \mathbf{S}$, $\widetilde{\nabla \cdot \mathbf{W}} = \frac{k^{5/2}}{\varepsilon^2}  \nabla \cdot \mathbf{W}$ and $\widetilde{\nabla k} = \frac{k^{1/2}}{\varepsilon}  \nabla k$ are the dimensionless counterparts of the symmetric part of the velocity gradient $\mathbf{S}$, the antisymmetric part of the velocity gradient $\mathbf{W}$, the divergence of $\mathbf{S}$, the divergence of $\mathbf{W}$  and the gradient of $k$,  respectively.

The list of dependences \eqref{eq:divtau_const_hp_not_final} can be simplified because $\nabla \cdot\mathbf{S} = \nabla \cdot\mathbf{W} = \frac{1}{2} \Delta \mathbf{u}$ from the Schwarz theorem and the incompressibility assumption. Indeed, if $\mathbf{u}$ is sufficiently regular, it holds
\begin{equation}
\begin{split}
&\nabla \cdot\mathbf{S} = \frac{\partial}{\partial x_j} \frac{1}{2} \Big[ \frac{\partial u_i}{\partial x_j} + \frac{\partial u_j}{\partial x_i} \Big] = \frac{1}{2} \Big[ \frac{\partial^2 u_i}{\partial x_j^2} + \frac{\partial}{\partial x_i} \frac{\partial u_j}{\partial x_j} \Big] =  \frac{1}{2} \frac{\partial^2 u_i}{\partial x_j^2} \\
& \nabla \cdot\mathbf{W} = \frac{\partial}{\partial x_j} \frac{1}{2} \Big[ \frac{\partial u_i}{\partial x_j} - \frac{\partial u_j}{\partial x_i} \Big] = \frac{1}{2} \Big[ \frac{\partial^2 u_i}{\partial x_j^2} - \frac{\partial}{\partial x_i} \frac{\partial u_j}{\partial x_j} \Big] =  \frac{1}{2} \frac{\partial^2 u_i}{\partial x_j^2}.
\end{split}
\end{equation}
As a consequence, the constitutive assumption can be simplified to 
\begin{equation} \label{eq:divtau_const_hp_not_final_1}
\widetilde{\divtau} = \mathbf{f}(\mathbf{s},\mathbf{w},\widetilde{\nabla \cdot \mathbf{S}},\widetilde{\nabla k}).
\end{equation}
In this work, a dependence from $\nabla \cdot \mathbf{S}$ and not from $\nabla \cdot \mathbf{s}$ (both to be made dimentionless) is supposed to make the implicit treatment of the first expansion term straightforwardly.

Finally, as remarked in \cite{Wu2018,Milani2020}, any other scalar quantity can be included in the constitutive assumption without changing the coordinates-frame rotation property. In particular, in our work we assume an additional dependence from the wall-distance based Reynolds number $Re_d$.

\subsection*{Implicit-Explicit treatment in \openfoam} 
The system \eqref{eq:Implicit_Explicit_RANS} does not require any coupling with a turbulence model and can theoretically be solved in \openfoam with a laminar solver like \texttt{icofoam}. The explicit term $(\divtau)^{\dag}$ is easy to implement because it is sufficient to define a new solver starting from an existing one by adding a constant source term into the momentum equation. 

The implicit term is less trivial to implement. The field $\nu^+_{tl}$ cannot be defined as a uniform field as the kinematic  viscosity. Therefore, it has been decided to implement a "fake" turbulence model that passes the same $\nu^+_{tl}$ field at each solver iteration. Thus, the \texttt{simpleFoam} solver is used. To the best knowledge of the authors, \openfoam is coded to deal with classic turbulent viscosities that are inside the divergence operator. In order to modify the code as less as possible, it has been decided to solve for an equivalent momentum equation that reads
\begin{equation} \label{eq:UEq_openfoam}
\frac{\partial \vel}{\partial t}+\vel \cdot \nabla\vel -  \nabla \cdot [(\nu+\nu^+_{tl}) \nabla \vel] = - \nabla p - (\divtau)^{\dag} - (\nabla \nu^+_{tl})^T \nabla \vel.
\end{equation}

%
%

\vskip1cm

\bibliographystyle{ieeetr} 
\bibliography{Bib_data_driven_div_tau}
\end{document}